\documentclass[aps, prl, twocolumn,10pt,showpacs,superscriptaddress,amsmath,amssymb,nobibnotes]{revtex4-1}

\usepackage{graphicx}
\usepackage{amsfonts}    
\usepackage{graphicx}   
\usepackage{verbatim}   
\usepackage{color}      
\usepackage{subfigure}  
\usepackage{hyperref}   
\usepackage{natbib}
\usepackage{booktabs}
\usepackage{threeparttable}
\usepackage{dcolumn}
\usepackage{multirow}

\begin{document}

\title{Observation of Nonlocality Sharing among Three Observers with One Entangled Pair via Optimal Weak Measurement}

\author{Meng-Jun Hu}\email{These authors contribute equally to this work}
\author{Zhi-Yuan Zhou}\email{These authors contribute equally to this work}
\author{Xiao-Min Hu}
\author{Chuan-Feng Li}
\author{Guang-Can Guo}
\author{Yong-Sheng Zhang}\email{yshzhang@ustc.edu.cn}
\affiliation{Laboratory of Quantum Information, University of Science and Technology of China, Hefei 230026, China}
\affiliation{CAS Center for Excellence in Quantum Information and Quantum Physics, University of Science and Technology of China, Hefei 230026, China}

\date{\today}

\begin{abstract}
\noindent Bell nonlocality plays a fundamental role in quantum theory. Numerous tests of the Bell inequality have been reported since the ground-breaking discovery of the Bell theorem. Up to now, however, most discussions of the Bell scenario have focused on a single pair of entangled particles distributed to only two separated observers. Recently,  it has been shown surprisingly that multiple observers can share the nonlocality from an entangled pair using the method of weak measurement without post-selection [Phys. Rev. Lett. {\bf 114}, 250401 (2015)]. Here we report an observation of double CHSH-Bell inequality violations for a single pair of entangled photons with strength continuous-tunable optimal weak measurement in a photonic system. Our results shed new light on the interplay between nonlocality and quantum measurements and our design of weak measurement protocol may also be significant for important applications such as unbounded randomness certification and quantum steering. 
\end{abstract}

\maketitle

\noindent{\bf INTRODUCTION} 

\noindent Nonlocality, which was pointed out by Einstein, Podolsky and Rosen (EPR) \cite{EPR}, plays a fundamental role in quantum theory. It has been intensively investigated since the ground-breaking discovery of Bell theorem by John Bell in 1964 \cite{Bell}. Bell theorem states that any local-realistic theory can not reproduce all the predictions of quantum theory and gives an experimental testable inequality \cite{speakable bell} that later improved by Clauser, Horne, Shimony and Holt (CHSH) \cite{CHSH}. 
Numerous tests of CHSH-Bell inequality have been realized in various quantum systems \cite{Clauser, Aspect, Zeilinger, Rowe, Monroe, Ansmann, Hofman, Giustina, Christensen} and strong loophole-free Bell tests have been reported recently \cite{Hensen, MGiustina, Shalm}.

To date, however, most discussions of Bell scenario focus on one pair of entangled particles distributed to only two separated observers Alice and Bob \cite{bunner}. 
It is thus an important and fundamental question whether or not multiple observers can share the nonlocality from an entangled pair. Using the concept of weak measurement without post-selection, Silva {\it et al}. give a surprising positive answer to above question and show a marvelous physical fact that measurement disturbance and information gain of a single system are closely related to nonlocality distribution among multiple observers in one entangled pair \cite{silva}.

In this article, we report an experimental realization of sharing nonlocality among three observers with strength continuous-tunable optimal weak measurement in a photonic system. We produce pairs of polarization entangled photons in our experiment and send it to Alice and Bob1, Bob2 separately, in which case Bobs access the same single particle from the entangled pairs with Bob1 performs weak measurement.
The realization of sharing nonlocality is certified by the observed double violations of CHSH-Bell inequality among Alice-Bob1 and Alice-Bob2. The reason behind this is because that weak measurement performed by Bob1 can be strong enough to obtain quantum correlations between Alice and Bob1, and weak enough to retain quantum correlations between Alice and Bob2.
Our results not only shed new light on the interplay between nonlocality and quantum measurements but also could find significant applications such as in unbounded randomness certification \cite{pironio, curchod} and quantum steering \cite{wiseman, vitus}.

\begin{figure}[bp]
\centering
\includegraphics[scale=0.4]{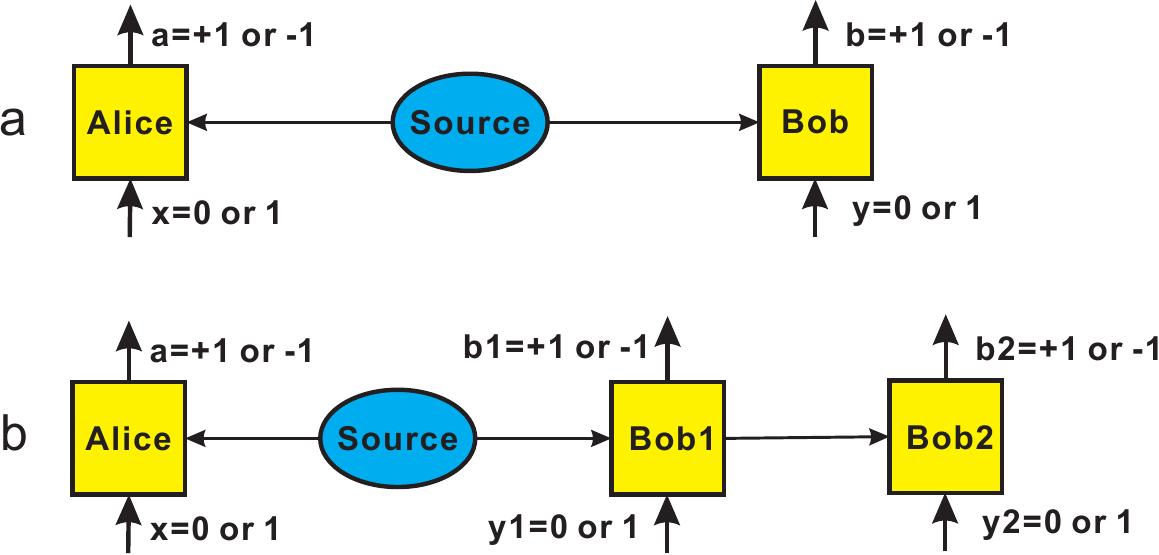}
\caption{Bell test.   a. Typical Bell scenario in which one pair of entangled particles is distributed to only two observers: Alice and Bob. b. Modified Bell scenario in which Bob1 and Bob2 access the same single particle from the entangled pair with Bob1 performs a weak measurement.}
\label{f1}
\end{figure}

\hfill

\noindent{\bf RESULTS}

\noindent As one of the foundations of quantum theory, the measurement postulate states that upon measurement, a quantum system will collapses into one of its eigenstates, with the probability determined by the Born rule. While this type of strong measurement, which is projective and irreversible, obtains the maximum information about a system, it also completely destroys the system after the measurement. 
Weak measurement, i.e., the coupling between the system and the probe is weak, however, can be used to extract less information about the system with smaller disturbance. It should be noted that this kind of weak disturbance measurement combined with post-selection usually refers to weak measurement\cite{Aharonov}, which has been shown to be a powerful method in signal amplification \cite{Hosten, Dixon, Xu}, state tomography \cite{Lundeen, lundeen} and in solving quantum paradoxes \cite{Botero} over the past decades. Hereafter, we follow the definition in \cite{silva} where weak measurement just refers to the measurement with intermediate coupling strength between the system and the probe. 
In contrast to strong projective measurement, weak measurement is non-destructive and retains some original properties of the measured system, e.g., coherence and entanglement. 
Because the entanglement is not completely destroyed by weak measurement, a particle that has been measured with intermediate strength can still be entangled with other particles, and therefore, shares nonlocality among multiple observers is possible.

Consider a von Neumann-type measurement \cite{von} on a spin-$1/2$ particle that is in the superposition state $|\psi\rangle=\alpha|\uparrow\rangle+\beta|\downarrow\rangle$ with $|\alpha|^{2}+|\beta|^{2}=1$ where $|\uparrow\rangle$ ($|\downarrow\rangle$) denotes the spin up (down) state. After the measurement, the spin state is entangled with the pointer's state, i.e., 
$|\psi\rangle\otimes|\phi\rangle\rightarrow\alpha|\uparrow\rangle\otimes|\phi_{\uparrow}\rangle+\beta|\downarrow\rangle\otimes|\phi_{\downarrow}\rangle$,
where $|\phi\rangle$ is the initial state of the pointer and $|\phi_{\uparrow}\rangle$ ($|\phi_{\downarrow}$) indicates the measurement result of spin up (down). By tracing out the state of the pointer, the spin state becomes
\begin{equation}
\rho=F\rho_{0}+(1-F)(\pi_{\uparrow}\rho_{0}\pi_{\uparrow}+\pi_{\downarrow}\rho_{0}\pi_{\downarrow}),
\end{equation}
where $\rho_{0}=|\psi\rangle\langle\psi|, \pi_{\uparrow}=|\uparrow\rangle\langle\uparrow|,\pi_{\downarrow}=|\downarrow\rangle\langle\downarrow|$ and $F=\langle\phi_{\downarrow}|\phi_{\uparrow}\rangle$. The quantity $F$ is called the measurement quality factor because it measures the disturbance of the measurement \cite{silva}. If $F=0$, the spin state is reduced to a completely decoherent state in the measurement eigenbasis, representing a strong measurement; otherwise, if $F=1$, there is no measurement at all. For other case, i.e., $F\in (0,1)$, it refers to the measurement with intermediate strength called weak measurement.

Another important quantity associated with weak measurement is the information gain $G$ that is determined by the precision of the measurement \cite{silva}. In the case of strong measurement, the probability of obtaining the outcome $+1$ ($-1$) that corresponds to spin eigenstate $|\uparrow\rangle$ ($|\downarrow\rangle$) can be calculated by the Born rule $P(+1)=\mathrm{Tr}(\pi_{\uparrow}\rho_{0})$ ($P(-1)=\mathrm{Tr}(\pi_{\downarrow}\rho_{0})$). However, the non-orthogonality of the pointer states $\langle\phi_{\uparrow}|\phi_{\downarrow}\rangle\neq 0$ in weak measurement results in ambiguous outcomes. An observer who performs a weak measurement must choose a complete orthogonal set of pointer states $\lbrace|\phi_{+1}\rangle, |\phi_{-1}\rangle\rbrace$ as reading states to define the outcomes $\lbrace +1, -1\rbrace $ corresponding to the spin eigenstates $\lbrace |\uparrow\rangle, |\downarrow\rangle\rbrace$.
The probabilities of the outcome $\pm1$ in weak measurement then become
$P(\pm1)=\mathrm{Tr}(\pi_{\uparrow}\rho_{0})|\langle\phi_{\pm1}|\phi_{\uparrow}\rangle|^{2}+\mathrm{Tr}(\pi_{\downarrow}\rho_{0})|\langle\phi_{\pm1}|\phi_{\downarrow}\rangle|^{2}$.
Here, $|\langle\phi_{+1}|\phi_{\uparrow}\rangle|^{2}$ and $|\langle\phi_{-1}|\phi_{\downarrow}\rangle|^{2}$ correspond to the probabilities of obtaining the correct outcomes while $|\langle\phi_{-1}|\phi_{\uparrow}\rangle|^{2}$ and $|\langle\phi_{+1}|\phi_{\downarrow}\rangle|^{2}$ correspond to the probabilities of the wrong outcomes.
For simplicity, we consider the case of symmetric ambiguousness in which $|\langle\phi_{+1}|\phi_{\uparrow}\rangle|^{2}=|\langle\phi_{-1}|\phi_{\downarrow}\rangle|^{2}$ and $|\langle\phi_{-1}|\phi_{\uparrow}\rangle|^{2}=|\langle\phi_{+1}|\phi_{\downarrow}\rangle|^{2}$, thus, the probabilities of the outcomes can be reformulated as
\begin{equation}
P(\pm1)=G\cdot\dfrac{1}{2}[1\pm\mathrm{Tr}(\sigma\rho_{0})]+(1-G)\cdot\dfrac{1}{2},
\end{equation}
where $\sigma=\pi_{\uparrow}-\pi_{\downarrow}$ defines the spin observable and $G=1-|\langle\phi_{-1}|\phi_{\uparrow}\rangle|^{2}-|\langle\phi_{+1}|\phi_{\downarrow}\rangle|^{2}$ represents the precision of the measurement (See more details in Methods). 
The quality factor $F$ and the precision $G$ are determined solely by the pointer states and satisfy the trade-off relation $F^{2}+G^{2}\leq1$ \cite{silva}. A weak measurement is optimal if $F^{2}+G^{2}=1$ is satisfied.
 
\begin{figure}[tbp]
\centering
\includegraphics[scale=0.45]{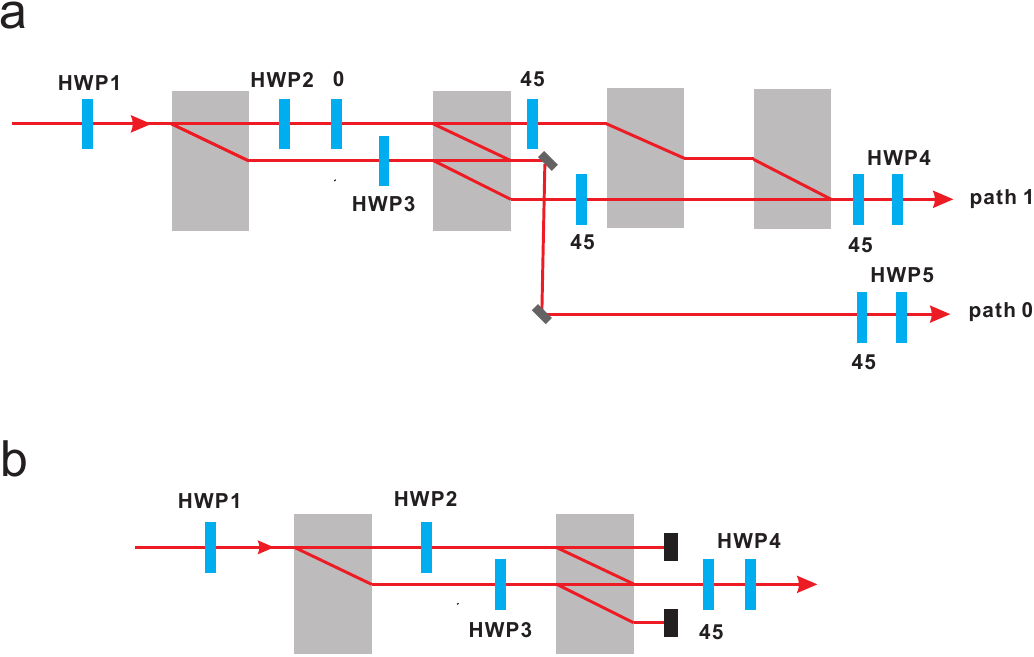}
\caption{ Optimal weak measurement realized in a photonic system. {\bf a:} HWP2 and HWP3 are rotated at $\theta/2$ and $\pi/4-\theta/2$ degree determining the strength of measurement $F=\mathrm{sin}2\theta$. Photons with vertical polarization state $|V\rangle$ transmit calcite beam displacer (BD) without change of its path while photons with horizontal polarization state $|H\rangle$ suffer a shift away from its original path. HWP1, HWP4 and HWP5 are rotated at the same degree $\varphi/2$ to realize weak measurement of polarization observable $\sigma_{\varphi}=|\varphi\rangle\langle\varphi|-|\varphi^{\perp}\rangle\langle\varphi^{\perp}|$. The measurement outcome $+1$($-1$) is encoded in path $0$($1$) separately. {\bf  b:} The setup, used in actual experiment, realizes same optimal weak measurement as shown above. The only difference is that specific outcome $+1$($-1$) can be selected by rotating HWP1 and HWP4. In the measurement of observation $\sigma_{\varphi}$ with HWP1 and HWP4 rotated at $\varphi/2$ degree, outcome $+1$ is obtained when photons comes out of the setup and outcome $-1$ is obtained when HWP1 and HWP4 rotated at $\varphi/2+\pi/4$. Note that measurement outcome values are extracted in the final coincidence detection. }
\label{f2}
\end{figure} 

\begin{figure*}[tbp]
\centering
\includegraphics[scale=0.9]{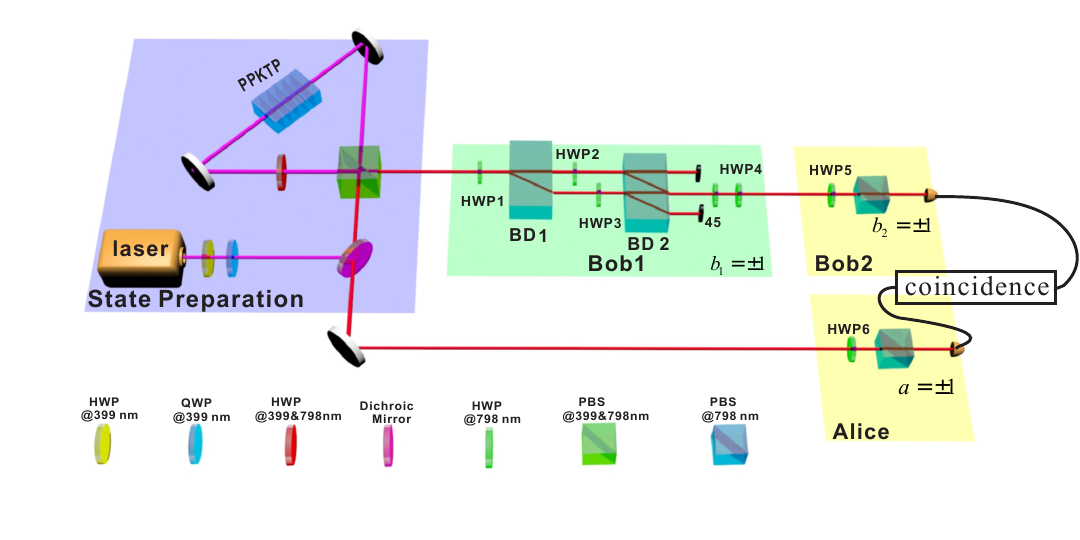}
\caption{ Measurement setup. Polarization-entangled pairs of photons are produced by pumping a type-II apodized  periodically poled potassium titanyl phosphate (PPKTP) crystal placed in the middle of a Sagnac-loop interferometer with dimensions of $1\mathrm{mm}\times 2\mathrm{mm}\times 20\mathrm{mm}$ and with end faces with anti-reflective coating at wavelengths of 399 nm and 798 nm.  
The photon emitted to Alice is measured via a combination of HWP6 and PBS. The green area shows the weak disturbance measurement setup of Bob1. During the experiment, HWP2, HWP3 are rotated by $\theta/2, \pi/4-\theta/2$ according to the experimental requirement. HWP1 is used for Bob1's measurement, and HWP4 is rotated by the same angle as HWP1 to transform the photons polarization state back to the measurement basis after the photon passes through two beam displacers (BDs). The photon passing through HWP4 is then sent to Bob2 for a strong projective measurement with HWP5 and PBS. In the final stage, two-photon coincidences at 6s are recorded by avalanche photodiode single-photon detectors and a coincidence counter (ID800).  }
\label{f3}
\end{figure*}

\hfill

\noindent{\bf Modified Bell test with weak measurement}
 
\noindent In a typical Bell test scenario, one pair of entangled spin-$1/2$ particles is distributed between two separated observers, Alice and Bob (Fig. \ref{f1}a), who each receive a binary input $x,y\in\lbrace0,1\rbrace$ and subsequently give a binary output $a,b\in\lbrace1,-1\rbrace$. For each input $x$ ($y$), Alice (Bob) performs a strong projective measurement of her (his) spin along a specific direction and obtains the outcome $a$ ($b$). 
The scenario is characterized by a joint probability distribution $P(ab|xy)$ of obtaining outcomes $a$ and $b$, conditioned on measurement inputs $x$ for Alice and $y$ for Bob. The fixed measurement inputs $x$ and $y$ defines the correlations
$C_{(x,y)}=\sum_{a,b}abP(ab|xy)$. 
The CHSH-Bell test is focused on the so-called $S$ value defined by the combination of correlations
\begin{equation}
S=|C_{(0,0)}+C_{(0,1)}+C_{(1,0)}-C_{(1,1)}|.
\end{equation}
While $S\leq 2$ in any local hidden variable theory \cite{CHSH}, quantum theory gives a more relaxed bound of $2\sqrt{2}$ \cite{observable}.

Here, we consider a new Bell scenario in which there are two observers Bob1 and Bob2 access to the same one-half of the entangled state of spin-$1/2$ particles (Fig. \ref{f1}b). 
Alice, Bob1 and Bob2 each receive a binary input $x,y_{1},y_{2}\in\lbrace0,1\rbrace$ and subsequently provide a binary output $a,b_{1},b_{2}\in\lbrace1,-1\rbrace$. For each input $y_{1}$, Bob1 performs weak measurement of his spin along a specific direction while Alice and Bob2 perform strong projective measurements for their input $x$ and $y_{2}$. With the outcome $b_{1}$, Bob1 sends the measured spin particle to Bob2. 
It should be emphasized here that the outcomes of Bob1 are actually obtained by Bob2 in our photonic experiment. This is because that the measurement of Bob1 is realized by coupling polarization of photons to its path and the outcomes are encoded in the path after measurement. 
The scenario is now characterized by joint conditional probabilities $P(ab_{1}b_{2}|xy_{1}y_{2})$, and an incisive question is raised whether Bob1 and Bob2 can both share nonlocality with Alice. 
The answer is surprisingly positive that the statistics of both Alice-Bob1 and Alice-Bob2 can indeed violate the CHSH-Bell inequality simultaneously \cite{silva}.

The quantities $G$ and $F$ of weak measurement, respectively, determine the $S$ values of Alice-Bob1 and Alice-Bob2 in the new Bell scenario. 
In the case that the Tsirelson's bound  $2\sqrt{2}$ of the CHSH-Bell inequality can be attained, the calculation gives (See more details in Methods)
\begin{equation}
S_{A-B1}=2\sqrt{2}G,   S_{A-B2}=\sqrt{2}(1+F).
\end{equation}

\hfill

\noindent{\bf Realization of optimal weak measurement in a photonic system}
 
\noindent To observe significant double violations of the CHSH-Bell inequality, the realization of optimal weak measurement is a key and necessary requirement. In the original scheme proposed in Ref. [18], the spatial degree of freedom of particle is used as the pointer. However, the particle with Gaussian spatial distribution only realizes sub-optimal weak measurement, i.e., $F^{2}+G^{2}< 1$ in their protocol.
Here, we propose and realize optimal weak measurement in a photonic system by using discrete pointer, i.e., path degree of freedom of photons instead of continuous pointer \cite{pryde}. It should be noted here that whether or not the pointer is continuous or discrete do not change any results discussed above.

Before illustration of the experimental realization, it should be emphasized first that weak measurement is mathematically equivalent to positive operator valued measures (POVMs) formalism \cite{mal} and this becomes our basis of experimental design. For the spin system discussed above, if Bob1 performs weak measurement and obtains outcome $\pm 1$, the states of measured system will accordingly collapse into
\begin{equation}
 |\Psi_{\pm 1}\rangle_{s}=\alpha\langle\phi_{\pm 1}|\phi_{\uparrow}\rangle|\uparrow\rangle+\beta\langle\phi_{\pm 1}|\phi_{\downarrow}\rangle|\downarrow\rangle
\end{equation} 
with probability $P(\pm 1)=\mathrm{Tr}(|\Psi_{\pm 1}\rangle_{s}\langle\Psi_{\pm 1}|)$. The weak measurement of Bob1 is actually to realize a two-outcome POVMs with Kraus operators \cite{Kraus}  
\begin{equation}
M_{\pm 1}=\langle\phi_{\pm 1}|\phi_{\uparrow}\rangle|\uparrow\rangle\langle\uparrow|+\langle\phi_{\pm 1}|\phi_{\downarrow}\rangle|\downarrow\rangle\langle\downarrow|
\end{equation}
corresponding to outcome $\pm 1$. 

In our realization of weak measurement of Bob1 with photonic elements as shown in Fig. \ref{f2}a, the measured photons are in polarization state and the path degree of freedom of photons is used as pointer. In order to perform weak measurement in specific polarization basis $\lbrace |\varphi\rangle, |\varphi^{\perp}\rangle\rbrace$ with defined observable $\sigma_{\varphi}=|\varphi\rangle\langle\varphi|-|\varphi^{\perp}\rangle\langle\varphi^{\perp}|$, we first transform the measured basis $\lbrace |\varphi\rangle, |\varphi^{\perp}\rangle\rbrace$ to basis $\lbrace |H\rangle, |V\rangle\rbrace$ via half wave plate (HWP1), then realize weak measurement of observable $\sigma_{H}=|H\rangle\langle H|-|V\rangle\langle V|$ via optical elements between HWP1 and HWP4, HWP5 and finally transform back to $\lbrace |\varphi\rangle, |\varphi^{\perp}\rangle\rbrace$ basis via HWP4 and HWP5. HWP1, HWP4 and HWP5 are rotated by the same angle $\varphi/2$.

The key part of our setup is the realization of weak measurement of observable $\sigma_{H}$ and this is achieved by interference between calcite beam displacers (BDs) (Fig. \ref{f2}). Consider photons with polarization state $|\Phi\rangle=\alpha|H\rangle+\beta|V\rangle$ to be measured, after interaction, the composite state of photons becomes $|\psi\rangle=\alpha|H\rangle|\phi_{H}\rangle+\beta|V\rangle|\phi_{V}\rangle$ with $|\phi_{H}\rangle$ ($|\phi_{V}\rangle$) is the corresponding pointer state. The reading states $\lbrace |\phi_{+1}\rangle, |\phi_{-1}\rangle\rbrace$ in our realization are chosen as states of two separated paths $0$ and $1$ (see Fig. 2a) denoted by $|0\rangle$ and $|1\rangle$. By rotating HWP2 and HWP3 between BDs at $\theta/2$ and $\pi/4-\theta/2$ degrees respectively, the pointer states become
\begin{equation}
\begin{split}
|\phi_{H}\rangle&=\mathrm{cos}\theta|0\rangle+\mathrm{sin}\theta|1\rangle, \\
|\phi_{V}\rangle&=\mathrm{sin}\theta|0\rangle+\mathrm{cos}\theta|1\rangle
\end{split}
\end{equation}
with $0\leq \theta\leq \pi/2$. The quality factor and information gain in our case are $F=\langle\phi_{H}|\phi_{V}\rangle=\mathrm{sin}2\theta$ and $G=1-|\langle 1|\phi_{H}\rangle|^{2}-|\langle 0|\phi_{V}\rangle|^{2}=\mathrm{cos}2\theta$. The condition of optimal weak measurement $F^{2}+G^{2}=1$ is satisfied. 

In practical experiment, we use the setup shown in Fig. \ref{f2}b instead of that shown in Fig. \ref{f2}a. The setup shown in Fig. 2b can realize the same optimal weak measurement as that in Fig. \ref{f2}a and the only difference is that specific outcome can be selected by rotating HWP1 and HWP4.  When Bob1 performs weak measurement of observable $\sigma_{\varphi}$ with HWP1 and HWP4 rotated at $\varphi/2$ (or $\pi/4-\varphi/2$) degree, photons comes out of setup have state $|\Psi_{+1}\rangle=M_{+1}|\Phi\rangle$ (or $|\Psi_{-1}\rangle=M_{-1}|\Phi\rangle$) corresponding to outcome $+1$ (or $-1$). Here, $M_{+1}=\mathrm{cos}\theta|\varphi\rangle\langle\varphi|-\mathrm{sin}\theta|\varphi^{\perp}\rangle\langle\varphi^{\perp}|$, $M_{-1}=\mathrm{sin\theta}|\varphi\rangle\langle\varphi|-\mathrm{cos}\theta|\varphi^{\perp}\rangle\langle\varphi^{\perp}|$ are Kraus operators and Bob1 extracts his measurement outcomes by final coincidence detection given that the rotation angles of HWP1 and HWP4 are known to him.

\begin{figure}[tbp]
\centering
\includegraphics[scale=0.33]{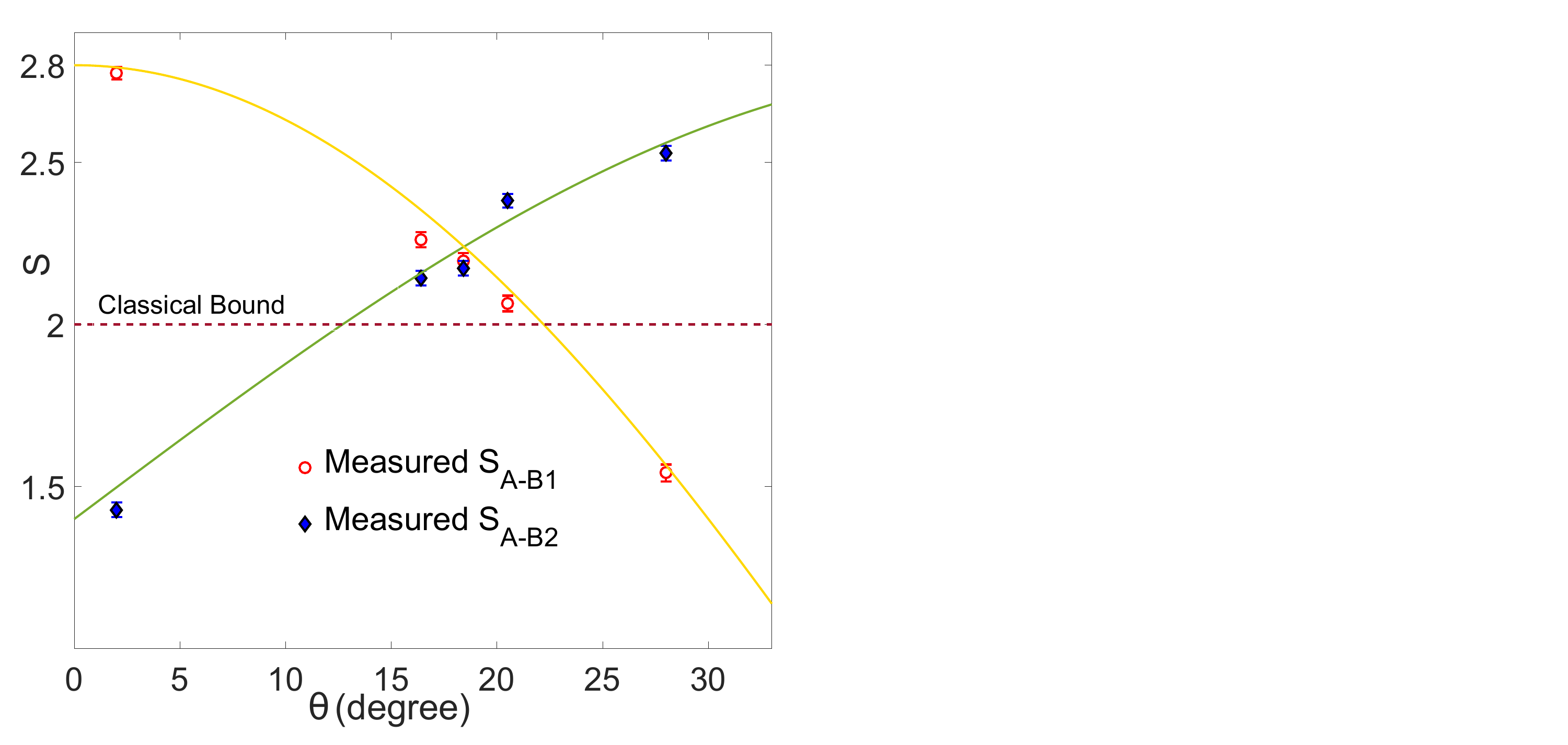}
\caption{ Experimental results. The yellow and green curves represent the theoretical predictions for $S_{A-B1}$ and $S_{A-B2}$, respectively, while the red circles and blue rhombus indicate the practical measured results for $S_{A-B1}$ and $S_{A-B2}$, corresponding $\theta=\lbrace 4^{\circ}, 16.4^{\circ}, 18.4^{\circ}, 20.5^{\circ}, 28^{\circ}\rbrace$. Double violations are observed in $\theta=\lbrace 16.4^{\circ}, 18.4^{\circ}, 20.5^{\circ}\rbrace$ with approximately $10$ standard deviations. The error bars are calculated according to Poissonian counting statistics. }
\label{f4}
\end{figure} 

\hfill

\noindent{\bf Experimental observation of double Bell inequality violations}
 
\noindent In our Bell test experiment (Fig. \ref{f3}), polarization-entangled pairs of photons in state $(|H\rangle|V\rangle-|V\rangle|H\rangle)/\sqrt{2}$ are generated by pumping a type-II apodized  periodically poled potassium titanyl phosphate (PPKTP) crystal to produce photon pairs at a wavelength of 798nm. A $4.5\mathrm{mW}$ pump laser centred at a wavelength of 399nm is produced by a Moglabs ECD004 laser, and a PPKTP crystal is embedded in the middle of a Sagnac interferometer to ensure the production of high-quality, high-brightness entangled pair \cite{new, zhou}. The maximum coincidence counting rates in the horizontal/vertical basis are approximately 3,200/s. The visibility of coincidence detection for the maximally entangled state is measured to be $0.997\pm 0.006$ in the horizontal/vertical polarization basis $\lbrace|H\rangle,|V\rangle\rbrace$ and $0.993\pm 0.008$ in the diagonal/antidiagonal polarization basis $\lbrace(|H\rangle\pm|V\rangle)/\sqrt{2}\rbrace$, achieved by rotating the polarization analyzers for two photons.

Alice, Bob1 and Bob2 each have two measurement choices, and for each choice, two trials are needed, corresponding to two different outcomes. For each fixed $\theta$, which determines the strength of the weak measurement $F=\mathrm{sin}2\theta$, we have implemented $64$ trials for calculating $S_{A-B1}$ and $S_{A-B2}$. To ensure that the Tsirelson's bound $2\sqrt{2}$ can be approached, Alice chooses measurement along direction $Z$ or $X$, while Bobs choose measurement along $(-Z+X)/\sqrt{2}$ or $-(Z+X)/\sqrt{2}$ direction. In this experiment, HWP6 is set at $(0^{\circ}, 45^{\circ})$ or $(22.5^{\circ}, 67.5^{\circ})$, corresponding to Alice's measurement along the $Z$ or $X$ direction, while HWP1 and HWP5, representing measurements of Bob1 and Bob2, are set at $(-11.25^{\circ}, 33.75^{\circ})$ or $(11.25^{\circ}, 56.25^{\circ})$, corresponding to the $(-Z+X)/\sqrt{2}$ or $-(Z+X)/\sqrt{2}$ direction, respectively. For instance, if HWP1, HWP4 and HWP5 are rotated at $-11.25^{\circ}$ and HWP6 is fixed at $0^{\circ}$, the three-variable joint conditional probability $P[a=1,b_{1}=1,b_{2}=1|x=Z,y_{1}=(-Z+X)/\sqrt{2},y_{2}=(-Z+X)/\sqrt{2}]$ is obtained by the final coincidence detection. The other joint conditional probabilities can be detected via similar various combination of HWP1, HWP4, HWP5 and HWP6.

Five different angles $\theta=\lbrace 4^{\circ}, 16.4^{\circ}, 18.4^{\circ}, 20.5^{\circ}, 28^{\circ}\rbrace$ are chosen from which the values of $\theta=\lbrace 16.4^{\circ}, 18.4^{\circ}, 20.5^{\circ}\rbrace$ are located in the region where double violations are predicted to be observed.
In particular, the balanced double violations $S_{A-B1}=S_{A-B2}=2.26$ are presented under optimal weak measurement when $F=0.6$, corresponding to $\theta=18.4^{\circ}$. 
Our final results are shown in Fig. \ref{f4}, where double violations are clearly displayed at $\theta=\lbrace 16.4^{\circ}, 18.4^{\circ}, 20.5^{\circ}\rbrace$ with approximately 10 standard deviations. Specifically, when $\theta=18.4^{\circ}$ we obtain $S_{A-B1}=2.20\pm 0.02$ and $S_{A-B2}=2.17\pm 0.02$.
Considering the possible statistical error, systematic error and imperfection of our apparatus, these experimental results fit well within the theoretical predictions.

\hfill

\noindent{\bf DISCUSSION}
 
\noindent In conclusion, we have observed double violations of the CHSH-Bell inequality for the entangled state of photon pairs by using a strength continuous-tunable optimal weak measurement. Our experimental results verify the nonlocality distribution among multiple observers and shed new light on our understanding of the fascinating properties of nonlocality and quantum measurement. 
The weak measurement technique used herein can find significant applications in unbounded randomness certification \cite{pironio, curchod} , which is a valuable resource applied from quantum cryptography \cite{hugo, scarani} and quantum gambling \cite{goldenberg, zhang} to quantum simulation \cite{simulation}.
Here, the $S$ value of the correlation between Alice and Bob2 is determined by the quality factor of Bob1's weak measurement, implying that Bob1 can control the nonlocal correlation of Alice and Bob2 by manipulating the strength of his measurement. 
This result provides tremendous motivation for the further quantum steering research \cite{wiseman, vitus}.

\noindent{\it Note added.} After we have submitted this work we noticed that double Bell inequality violations was also observed using similar method in Ref. \cite{similar} though they did not realize optimal weak measurement.

\hfill

\noindent{\bf METHODS}

\noindent{\bf Weak measurement on a spin-1/2 particles}

\noindent Consider a typical measurement of spin observable $\sigma_{z}$ of a spin-1/2 particle with eigenstates satisfy $\sigma_{z}|\uparrow\rangle=|\uparrow\rangle$ and $\sigma_{z}|\downarrow\rangle=-|\downarrow\rangle$. The initial pointer state $|\phi\rangle$ is entangled with spin system after measurement interaction
\begin{equation}
\begin{split}
\hat{U}(\varrho\otimes|\phi\rangle\langle\phi|)\hat{U^{\dagger}}&=\pi_{\uparrow}\varrho\pi_{\uparrow}\otimes|\phi_{\uparrow}\rangle\langle\phi_{\uparrow}|+\pi_{\downarrow}\varrho\pi_{\downarrow}\otimes|\phi_{\downarrow}\rangle\langle\phi_{\downarrow}| \\
&+\pi_{\uparrow}\varrho\pi_{\downarrow}\otimes|\phi_{\uparrow}\rangle\langle\phi_{\downarrow}|+\pi_{\downarrow}\varrho\pi_{\uparrow}\otimes|\phi_{\downarrow}\rangle\langle\phi_{\uparrow}|,
\end{split}
\end{equation}
where $\varrho$ represents the initial state of spin system, $\pi_{\uparrow}\equiv|\uparrow\rangle\langle\uparrow|,\pi_{\downarrow}\equiv|\downarrow\rangle\langle\downarrow|$ are the projectors on the eigenstates of $\sigma_{z}$ and $|\phi_{\uparrow}\rangle,|\phi_{\downarrow}\rangle$ are the evolved pointer states corresponding to $|\uparrow\rangle,|\downarrow\rangle$ respectively.

The state of spin system, after tracing out pointer, becomes
\begin{equation}
\rho=\pi_{\uparrow}\varrho\pi_{\uparrow}+\pi_{\downarrow}\varrho\pi_{\downarrow}+\pi_{\uparrow}\varrho\pi_{\downarrow}\otimes\langle\phi_{\downarrow}|\phi_{\uparrow}\rangle+\pi_{\downarrow}\varrho\pi_{\uparrow}\otimes\langle\phi_{\uparrow}|\phi_{\downarrow}\rangle.
\end{equation}
To quantify the disturbance of measurement to the spin system, a quantity $F$ called {\it quality factor} of measurement can be defined \cite{silva}
\begin{equation}
F\equiv\langle\phi_{\uparrow}|\phi_{\downarrow}\rangle.
\end{equation}
Usually, $F$ is a complex value. Without loss of generality, here we take it as a real number for the simplicity of discussion. The Eq. (9) thus can be reformulated as
\begin{equation}
\rho=F\varrho+(1-F)(\pi_{\uparrow}\varrho\pi_{\uparrow}+\pi_{\downarrow}\varrho\pi_{\downarrow})
\end{equation}
since that $\varrho=\pi_{\uparrow}\varrho\pi_{\uparrow}+\pi_{\downarrow}\varrho\pi_{\downarrow}+\pi_{\uparrow}\varrho\pi_{\downarrow}+\pi_{\downarrow}\varrho\pi_{\uparrow}$.

If $F=0$, the spin up state $|\uparrow\rangle$ and spin down state $|\downarrow\rangle$ can be distinguished definitely through the orthogonal pointer states $|\phi_{\uparrow}\rangle$ and $|\phi_{\downarrow}\rangle$. The state of spin system $\rho$ is thus reduced to a state of completely decohered in the eigenbasis of $\sigma_{z}$ and the measurement in this case is called a strong measurement in which we can obtain the maximum information about the system. There is no measurement at all if $F=1$, i.e., the pointer state is the same for spin up or spin down. The measurement is called weak when $F\in(0,1)$ in which we can obtain partial information of the spin states with partial disturbance on it. It is obvious from Eq. (11) that a weak measurement can be considered as the combination of a strong measurement and none of measurement operationally. The quality factor $F$ thus reflects the strength of measurement and the disturbance of measurement.

The measurement of a spin-1/2 system only have two outcomes $+1$ and $-1$ corresponding to eigenstates $|\uparrow\rangle$ and $|\downarrow\rangle$ respectively. If the measurement is strong, the probability of obtaining outcome $+1$ (or $-1$) can be easily determined by the Born rule that $P(+1)=\mathrm{Tr}(\pi_{\uparrow}\varrho)$ (or $P(-1)=\mathrm{Tr}(\pi_{\downarrow}\varrho)$). However, in weak measurement the nonorthogonality of pointer states $F=\langle\phi_{\uparrow}|\phi_{\downarrow}\rangle\neq 0$ brings the uncertainty ambiguity.
In practice, an observer, who performs weak measurement, has to choose a complete orthogonal set of pointer states ${|\phi_{+1}\rangle,|\phi_{-1}\rangle}$ as reading states of pointer to define outcomes ${+1,-1}$ corresponding to the spin eigenstates ${|\uparrow\rangle,|\downarrow\rangle}$. 
To calculate the probability of outcome $+1$ (or $-1$), the state of pointer after measurement have to be considered. Tracing out the spin degree of freedom in Eq.  (8), we obtain the post-measurement state of the pointer
\begin{equation}
\rho_{p}=\mathrm{Tr}(\pi_{\uparrow}\varrho)\otimes|\phi_{\uparrow}\rangle\langle\phi_{\uparrow}|+\mathrm{Tr}(\pi_{\downarrow}\varrho)\otimes|\phi_{\downarrow}\rangle\langle\phi_{\downarrow}|.
\end{equation}
The probability of outcome $+1$ (or $-1$) now becomes the probability of obtaining the state $|\phi_{+1}\rangle$ (or $|\phi_{-1}\rangle$) of pointer. The probability of outcome $\pm1$ becomes
\begin{equation}
P(\pm1)=\mathrm{Tr}(\pi_{\uparrow}\varrho)|\langle\phi_{\pm1}|\phi_{\uparrow}\rangle|^{2}+\mathrm{Tr}(\pi_{\downarrow}\varrho)|\langle\phi_{\pm1}|\phi_{\downarrow}\rangle|^{2}.
\end{equation}

If the pointer states are considered in position representation, the observer can associate positive positions to outcome $+1$ and negative positions to outcome $-1$. In general, the reading states of pointer ${|\phi_{+1}\rangle,|\phi_{-1}\rangle}$ should be chosen that the maximum of $|\langle\phi_{+1}|\phi_{\uparrow}\rangle|^{2}$ and $|\langle\phi_{-1}|\phi_{\downarrow}\rangle|^{2}$ are achieved. If the reading states ensure that $|\langle\phi_{+1}|\phi_{\uparrow}\rangle|^{2}=|\langle\phi_{-1}|\phi_{\downarrow}\rangle|^{2}$, i.e., the probabilities of obtaining correct outcomes are equal for spin up and spin down, the measurement is called unbiased. Here we only focus on unbiased measurement and assume that the reading states satisfy conditions $|\langle\phi_{+1}|\phi_{\uparrow}\rangle|^{2}=|\langle\phi_{-1}|\phi_{\downarrow}\rangle|^{2}$ and $|\langle\phi_{+1}|\phi_{\downarrow}\rangle|^{2}=|\langle\phi_{-1}|\phi_{\uparrow}\rangle|^{2}$.   In this case, the probability $P(\pm1)$ can be reformulated as
\begin{equation}
P(\pm1)=G\cdot\dfrac{1}{2}[1\pm\mathrm{Tr}(\sigma_{z}\varrho)]+(1-G)\cdot\dfrac{1}{2},
\end{equation}
where $\sigma_{z}=\pi_{\uparrow}-\pi_{\downarrow}$ is the spin observable and $G=1-|\langle\phi_{+1}|\phi_{\downarrow}\rangle|^{2}-|\langle\phi_{-1}|\phi_{\uparrow}\rangle|^{2}=1-2|\langle\phi_{+1}|\phi_{\downarrow}\rangle|^{2}$ defines the precision of measurement.
The precision of measurement $G$ indicates the correctness or the extent of unambiguity of outcomes.

The state of the spin system, after the observer obtains outcome $\pm1$, becomes
\begin{equation}
\begin{split}
\rho_{\pm1}&=\langle\phi_{\pm1}|\hat{U}\varrho\otimes|\phi\rangle\langle\phi|\hat{U}^{\dagger}|\phi_{\pm1}\rangle \\
&=\pi_{\uparrow}\varrho\pi_{\uparrow}|\langle\phi_{\pm1}|\phi_{\uparrow}\rangle|^{2}+\pi_{\downarrow}\varrho\pi_{\downarrow}|\langle\phi_{\pm1}|\phi_{\downarrow}\rangle|^{2} \\
&+\pi_{\uparrow}\varrho\pi_{\downarrow}\langle\phi_{\pm1}|\phi_{\uparrow}\rangle\langle\phi_{\downarrow}|\phi_{\pm1}\rangle+\pi_{\downarrow}\varrho\pi_{\uparrow}\langle\phi_{\pm1}|\phi_{\downarrow}\rangle\langle\phi_{\uparrow}|\phi_{\pm1}\rangle.
\end{split}
\end{equation}
Since that $|\langle\phi_{+1}|\phi_{\uparrow}\rangle|^{2}=\dfrac{1+G}{2},|\langle\phi_{+1}|\phi_{\downarrow}\rangle|^{2}=\dfrac{1-G}{2}$ and $\langle\phi_{+1}|\phi_{\downarrow}\rangle\langle\phi_{\uparrow}|\phi_{+1}\rangle=\langle\phi_{+1}|\phi_{\uparrow}\rangle\langle\phi_{\downarrow}|\phi_{+1}\rangle=\dfrac{F}{2}$, Eq. (15) can be reformulated as (unnormalized)
\begin{equation}
\rho_{\pm1}=\dfrac{F}{2}\varrho+\dfrac{1\pm G-F}{2}\pi_{\uparrow}\varrho\pi_{\uparrow}+\dfrac{1\mp G-F}{2}\pi_{\downarrow}\varrho\pi_{\downarrow}
\end{equation}
The quality factor $F$ and the precision $G$ satisfy the trade-off relation \cite{silva} 
\begin{equation}
F^{2}+G^{2}\leq 1.
\end{equation}
A weak measurement is optimal if $F^{2}+G^{2}=1$, otherwise it is suboptimal.

\hfill

\noindent{\bf Relation between weak measurement and Bell nonlocality}

\noindent The connection between weak measurement and nonlocality can be shown in the new Bell scenario that one pair of entangled spin-$\dfrac{1}{2}$ particles are delivered to Alice, Bob1 and Bob2, here Bob1 and Bob2 access to the same particle shown in Fg. 1 of main text. Contrary to Bob2 who performs the strong measurement, Bob1 performs the weak measurement before Bob2. After the measurement of Bob1, the particle will be sent to Bob2 who has no idea of Bob1's existence. Suppose that the entangled pair is in the singlet state
\begin{equation}
|\Psi\rangle=\dfrac{1}{\sqrt{2}}(|\uparrow\rangle|\downarrow\rangle-|\downarrow\rangle|\uparrow\rangle).
\end{equation}
Similar to the standard Bell scenario, Alice, Bob1 and Bob2 each receives a binary input $x,y_{1},y_{2}\in\lbrace 0,1\rbrace$ and accordingly performs measurement of their spin along the corresponding direction $\vec{\lambda}_{x},\vec{\mu}_{y_{1}}, \vec{\nu}_{y_{2}}$ respectively. The outcomes of their measurement are labelled by $a,b_{1},b_{2}\in \lbrace +1,-1\rbrace$. To study correlations between Alice and Bobs, we need to calculate the conditional probability distributions $P(ab_{1}b_{2}|xy_{1}y_{2})$ that can be simplified by using no-signalling condition
\begin{equation}
P(ab_{1}b_{2}|xy_{1}y_{2})=P(a|x)P(b_{1}|xy_{1}a)P(b_{2}|xy_{1}y_{2}ab_{1}).
\end{equation}

The probability of obtaining outcome $a$ conditioned on $x$ for Alice can be easily shown to be $P(a|x)=\dfrac{1}{2}$ since that any strong measurement on one-half of singlet state gives outcomes with equal probability.  
After the measurement of Alice, the spin state of another particle sent to Bob1 will collapse into the state that in an opposite spin direction with respect to Alice's post-measurement state
\begin{equation}
\rho_{|xa}=\pi_{-a\vec{\lambda}_{x}}=\dfrac{1}{2}(I-a\vec{\lambda}_{x}\cdot\vec{\sigma}),
\end{equation}
where $\pi_{-a\vec{\lambda}_{x}}$ represents the spin projector along the direction $-a\vec{\lambda}_{x}$ and $I,  \vec{\sigma}$ are identity operator and Pauli operator respectively.
The measurement of Bob1 is weak, the probability $P(b_{1}|xy_{1}a)$ is determined by Eq. (14) and
\begin{equation}
\begin{split}
P(b_{1}|xy_{1}a)&=G\cdot\mathrm{Tr}(\pi_{b_{1}\vec{\mu}_{y_{1}}}\rho_{|xa})+(1-G)\cdot\dfrac{1}{2} \\
&=\dfrac{1-Gab_{1}\vec{\lambda}_{x}\cdot\vec{\mu}_{y_{1}}}{2},
\end{split}
\end{equation}
where $\mathrm{Tr}(\pi_{b_{1}\vec{\mu}_{y_{1}}}\rho_{|xa})=\dfrac{1-ab_{1}\vec{\lambda}_{x}\cdot\vec{\mu}_{y_{1}}}{2}$ and $G$ is the precision of Bob1's weak measurement.
The spin state of Bob1's particle after weak measurement, according to Eq. (16), becomes
\begin{equation}
\begin{split}
\rho_{|xy_{1}ab_{1}}&=\dfrac{F}{2}\rho_{|xa}+\dfrac{1+b_{1}G-F}{2}\pi_{\vec{\mu}_{y_{1}}}\rho_{|xa}\pi_{\vec{\mu}_{y_{1}}} \\
&+\dfrac{1-b_{1}G-F}{2}\pi_{-\vec{\mu}_{y_{1}}}\rho_{|xa}\pi_{-\vec{\mu}_{y_{1}}}
\end{split}
\end{equation} 
with its norm trace $\mathrm{Tr}(\rho_{|xy_{1}ab_{1}})=P(b_{1}|xy_{1}a)$ and $F$ is quality factor of the weak measurement. The probability of obtaining outcome $b_{2}$ for Bob2's strong measurement is 
\begin{equation}
\begin{split}
& P(b_{2}|xy_{1}y_{2}ab_{1})=\mathrm{Tr}(\pi_{b_{2}\vec{\nu}_{y_{2}}}\tilde{\rho}_{|xy_{1}ab_{1}})  \\
=&\dfrac{1}{P(b_{1}|xy_{1}a)}\lbrace\dfrac{F}{4}(1-ab_{2}\vec{\lambda}_{x}\cdot\vec{\nu}_{y_{2}}) \\
& +\dfrac{1-F}{4}[1-ab_{2}(\vec{\lambda}_{x}\cdot\vec{\mu}_{y_{1}})(\vec{\mu}_{y_{1}}\cdot\vec{\nu}_{y_{2}})] \\
& +\dfrac{b_{1}G}{4}(b_{2}\vec{\mu}_{y_{1}}\cdot\vec{\nu}_{y_{2}}-a\vec{\lambda}_{x}\cdot\vec{\mu}_{y_{1}})\rbrace,
\end{split}
\end{equation}
where $\tilde{\rho}_{|xy_{1}ab_{1}}=\dfrac{1}{P(b_{1}|xy_{1}a)}\rho_{|xy_{1}ab_{1}}$ is a normalized state. 

Now we can calculate conditional probabilities $P(ab_{1}b_{2}|xy_{1}y_{2})$ according to Eq. (12)
\begin{equation}
\begin{split}
P(ab_{1}b_{2}|xy_{1}y_{2})& =\dfrac{F}{4}(\dfrac{1-ab_{2}\vec{\lambda}_{x}\cdot\vec{\nu}_{y_{2}}}{2}) \\
& +\dfrac{1-F}{4}[\dfrac{1-ab_{2}(\vec{\lambda}_{x}\cdot\vec{\mu}_{y_{1}})(\vec{\mu}_{y_{1}}\cdot\vec{\nu}_{y_{2}})}{2}] \\
& -\dfrac{b_{1}G}{4}(\dfrac{a\vec{\lambda}_{x}\cdot\vec{\mu}_{y_{1}}-b_{2}\vec{\mu}_{y_{1}}\cdot\vec{\nu}_{y_{2}}}{2}).
\end{split}
\end{equation}
Since probability lies between $0$ and $1$, if $\vec{\lambda}_{0}=\vec{Z}, \vec{\mu}_{0}=-\vec{X}$ and $\vec{\nu}_{0}=\vec{Z}\mathrm{sin}\theta-\vec{X}\mathrm{cos}\theta$ are chosen, we obtain
\begin{equation}
P(1-11|000)=F\mathrm{sin}\theta+G\mathrm{cos}\theta\leq 1
\end{equation}
along with outcomes $a=b_{2}=1$ and $b_{1}=-1$. This is the expression of a tangent to the unit circle $F^{2}+G^{2}=1$ and obviously the optimal pointer saturates this constraint.

The nonlocal correlation of Alice and Bobs's can be shown by calculating $S$ value defined as
\begin{equation}
S_{Alice-Bob_{n}}=|C_{(0,0)}^{n}+C_{(0,1)}^{n}+C_{(1,0)}^{n}-C_{(1,1)}^{n}|,
\end{equation}
where $n\in\lbrace 1,2\rbrace$ and $C_{(x,y_{n})}^{n}$ defines correlation of Alice and Bobn's measurement outcomes
\begin{equation}
C_{(x,y_{n})}^{n}=\mathrm{Tr}(\rho_{n}\sigma_{\vec{\lambda}_{x}}\otimes\sigma_{\vec{\mu}_{y_{n}}})=\sum_{a,b_{n}}ab_{n}P(ab_{n}|xy_{n})
\end{equation}
with $\rho_{n}$ is the state of the spin-$\dfrac{1}{2}$ entangled pair possessed by Alice and Bobs, $\sigma_{\vec{\lambda}_{x}},\sigma_{\vec{\mu}_{y_{n}}}$ represent the spin observables corresponding to directions $\vec{\lambda}_{x}$ and $\vec{\mu}_{y_{n}}$ respectively.

Since that $P(ab_{1}|xy_{1})=P(a|x)P(b_{1}|xy_{1}a)=\dfrac{1-Gab_{1}\vec{\lambda}_{x}\cdot\vec{\mu}_{y_{1}}}{4}$, the correlation $C_{(x,y_{1})}$ of Alice and Bob1 is $C_{(x,y_{1})}=-G\vec{\lambda}_{x}\cdot\vec{\mu}_{y_{1}}$ and thus we have 
\begin{equation}
S_{A-B1}=G\cdot|\vec{\lambda}_{0}\cdot\vec{\mu}_{0}+\vec{\lambda}_{0}\cdot\vec{\mu}_{1}+\vec{\lambda}_{1}\cdot\vec{\mu}_{0}-\vec{\lambda}_{1}\cdot\vec{\mu}_{1}|.
\end{equation}
Similarly
\begin{equation}
\begin{split}
P(ab_{2}|xy_{2})&=\sum_{b_{1},y_{1}}P(ab_{1}b_{2}|xy_{1}y_{2}) \\
&=\dfrac{F}{4}(\dfrac{1-ab_{2}\vec{\lambda}_{x}\cdot\vec{\nu}_{y_{2}}}{2}) \\
& +\dfrac{1-F}{4}\sum_{y_{1}}[\dfrac{1-ab_{2}(\vec{\lambda}_{x}\cdot\vec{\mu}_{y_{1}})(\vec{\mu}_{y_{1}}\cdot\vec{\nu}_{y_{2}}}{2})]
\end{split}
\end{equation}
and
\begin{equation}
C_{(x,y_{2})}=-\dfrac{F}{2}\vec{\lambda}_{x}\cdot\vec{\nu}_{y_{2}}-\dfrac{1-F}{2}\sum_{y_{1}}(\vec{\lambda}_{x}\cdot\vec{\mu}_{y_{1}})(\vec{\mu}_{y_{1}}\cdot\vec{\nu}_{y_{2}}).
\end{equation}
The $S$ value of Alice and Bob2 is calculated as
\begin{equation}
\begin{split}
S_{A-B2}&=|\dfrac{F}{2}(\vec{\lambda}_{0}\cdot\vec{\nu}_{0}+\vec{\lambda}_{0}\cdot\vec{\nu}_{1}+\vec{\lambda}_{1}\cdot\vec{\nu}_{0}-\vec{\lambda}_{1}\cdot\vec{\nu}_{1}) \\
&-\dfrac{1-F}{2}\sum_{y_{1}}[(\vec{\lambda}_{0}\cdot\vec{\mu}_{y_{1}})(\vec{\mu}_{y_{1}}\cdot\vec{\nu}_{0})+(\vec{\lambda}_{0}\cdot\vec{\mu}_{y_{1}})(\vec{\mu}_{y_{1}}\cdot\vec{\nu}_{1}) \\
& +(\vec{\lambda}_{1}\cdot\vec{\mu}_{y_{1}})(\vec{\mu}_{y_{1}}\cdot\vec{\nu}_{0})-(\vec{\lambda}_{1}\cdot\vec{\mu}_{y_{1}})(\vec{\mu}_{y_{1}}\cdot\vec{\nu}_{1})]|.
\end{split}
\end{equation}

In the case that quantum bound $2\sqrt{2}$ is obtained, i.e., Alice measures in the directions $\vec{Z}$ or $\vec{X}$ according to her inputs $0$ or $1$, while Bob1 and Bob2 measure in the directions $\dfrac{-(\vec{Z}+\vec{X})}{\sqrt{2}}$ or $\dfrac{-\vec{Z}+\vec{X}}{\sqrt{2}}$ for their respective inputs $0$ or $1$, we obtain
\begin{equation}
\begin{split}
S_{A-B1}&=2\sqrt{2}G        \\
S_{A-B2}&=\sqrt{2}(1+F),
\end{split}
\end{equation}
which implies that the nonlocality correlation of Alice and Bob1 and correlation of Alice and Bob2 are totally determined by the weak measurement performed by Bob1. Double violations happen when quality factor $F$ and precision $G$ of weak measurement satisfy $F\in(\sqrt{2}-1,1]$ and $G\in(\dfrac{\sqrt{2}}{2},1]$. Restricted by the general condition $F^{2}+G^{2}\leq 1$, the area that double violations exist is limited. In the case that optimal weak measurement is realized, which means $F^{2}+G^{2}=1$, double violations can be observed only when $F\in(\sqrt{2}-1,\dfrac{\sqrt{2}}{2})$. If $F=0.6$ and $ G=0.8$ are chosen, we obtain the optimal double violations $S_{Alice-Bob1}=S_{Alice-Bob2}\approx 2.26$.

\hfill

\noindent{\bf Demonstration of entangled photons source in experiment}

\noindent In our experiment, high quality polarization entangled photon source is produced by pumping a type-II apodized periodically poled patassium titanyl phosphate (PPKTP) crystal inside a Sagnac-loop interferometer. The PPKTP has dimensions of $1\mathrm{mm}\times2\mathrm{mm}\times20\mathrm{mm}$ and the end faces are anti-reflective coated at wavelengths of $399\mathrm{nm}$ and $798\mathrm{nm}$. The temperature of the crystal is controlled by using a home-made temperature controller with stability of $\pm 2\mathrm{mK}$. The ultraviolet (UV) pump beam is generated from a commercial Moglabs ECD004 laser. One UV quarter wave plate (QWP) and one UV half wave plate (HWP) are placed at the input port of the interferometer for controlling the power and relative phase of pump beam inside the Sagnac-loop interferometer. Polarization orthogonal pump beams are separated by a dual wavelength polarized beam splitter (DPBS). The vertical polarized pump beam is rotated to horizontal polarization by using a dual wavelength HWP before interact with the PPKTP crystal for spontaneous parametric down-conversion (SPDC). Orthogonal polarized photon pairs are generated in two counter propagating directions combined at the DPBS. The photon emitted to Alice is first separated from the pump beam by using a dichromatic mirror (DM) and then measured projectively via combination of HWP and PBS by Alice. The photon sent to Bob, first passes through the weak measurement seup of Bob1 and is subsequently sent to Bob2 for projective measurement.

\begin{figure}[tbp]
\centering
\includegraphics[scale=0.28]{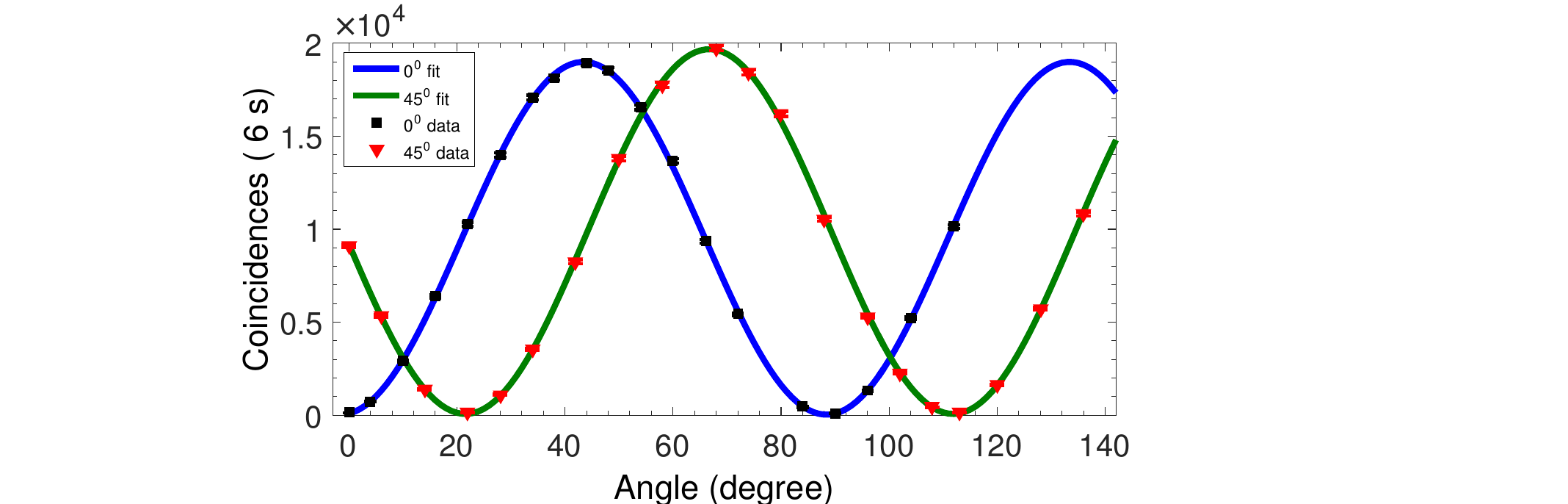}
\caption{Polarization interference results. }
\label{f5}
\end{figure}

The state of photon pair output from the interferometer can be expressed as
\begin{equation}
|\psi\rangle=\dfrac{1}{\sqrt{2}}(|H\rangle|V\rangle+e^{i\vartheta}|V\rangle|H\rangle),
\end{equation}
where the relative phase $\vartheta$ is determined by the relative position of QWP and HWP at the input port. In our experiment, the phase $\vartheta$ is tuned to $\pi$ such that the singlet state is produced
\begin{equation}
|\phi^{-}\rangle=\dfrac{1}{\sqrt{2}}(|H\rangle|V\rangle-|V\rangle|H\rangle).
\end{equation}

The quality of our source is characterized by using a two-photon polarization interference shown in Fig. \ref{f5}. The $399\mathrm{nm}$ wavelength pump beam's power is fixed at $4.5\mathrm{mW}$ and the coincidence windows is set in $2\mathrm{ns}$. The single counts in $6s$ is about $375000$ and $275000$ and the maximum coincidence is about $19000$. The raw visibilities in $0^{\circ}/90^{\circ}$ and $45^{\circ}/-45^{\circ}$ are $(99.70\pm 0.06)$ percents and $(99.32\pm 0.08)$ percents respectively. Therefore high visibilities guarantee the large violation of Bell-CHSH inequality.

\hfill

\noindent {\bf Data availability}

\noindent The data that support the findings of this study are available from the corresponding author upon reasonable
request.

\hfill

\noindent{\bf ACKNOWLEDGEMENTS}

\noindent The authors thank Yun-Feng Huang, Bi-Heng Liu and Bao-Sen Shi for helpful discussions and technical supports. Meng-Jun Hu also acknowledges Sheng Liu, Zhi-Bo Hou, Jian Wang, Chao Zhang, Ya Xiao, Dong-Sheng Ding, Yong-Nan Sun and Geng-Chen for many useful suggestions and stimulated discussions. This work was supported by the National Natural Science Foundation of China (No. 61275122, No. 11674306 and 61590932 ), the Strategic Priority Research Program (B) of the Chinese Academy of Sciences (No. XDB01030200) and National key R$\&$D program (No. 2016YFA0301300 and No. 2016YFA0301700).

\hfill

\noindent{\bf Competing interests}

\noindent The authors declare that they have no competing interests.

\hfill

\noindent{\bf AUTHOR CONTRIBUTIONS}

\noindent Yong-Sheng Zhang and Meng-Jun Hu start the project and design the experiment. Zhi-Yuan Zhou, Meng-Jun Hu and Xiao-Min Hu perform the experiment and complete the data analysis. Meng-Jun Hu, Chuan-Feng Li and Yong-Sheng Zhang provide the theoretical calculations. Yong-Sheng Zhang and Guang-Can Guo supervise the project. Meng-Jun Hu write the manuscript and all authors participate in discussions. Meng-Jun Hu and Zhi-Yuan Zhou contribute equally to this work.

\hfill

\end{document}